\documentclass[conference]{IEEEtran}
\IEEEoverridecommandlockouts

\usepackage{makecell}
\usepackage[dvips]{graphicx}
\usepackage{latexsym}
\usepackage{amssymb}
\usepackage{amsmath}
\usepackage{bm}
\usepackage{multirow}
\usepackage{xcolor}
\usepackage{lipsum}
\usepackage{multicol}
\usepackage{enumerate}
\usepackage{algorithm}
\usepackage{algorithmicx}
\usepackage{algpseudocode}
\usepackage{amsmath}
\usepackage{graphicx}
\usepackage{subfig}
\usepackage[utf8]{inputenc}
\usepackage{float}

\def\BibTeX{{\rm B\kern-.05em{\sc i\kern-.025em b}\kern-.08em
    T\kern-.1667em\lower.7ex\hbox{E}\kern-.125emX}}
\begin{document}

\title{Symbol-Level Precoding Made Practical for Multi-Level Modulations via Block-Level Rescaling
}
\author{\IEEEauthorblockN{Ang Li$^*$, Fan Liu$^\ddag$, Xuewen Liao$^*$, Yuanjun Shen$^\dag$, and Christos Masouros$^\dag$}
\IEEEauthorblockA{
School of Information and Communications Engineering, Xi'an Jiaotong University, Xi'an, China$^*$\\
Department of Electrical and Electronic Engineering, Southern University of Science and Technology, Shenzhen 518055, China$^\ddag$\\
Department of Electronic and Electrical Engineering, University College London, London, UK$^\dag$\\
Email: \{ang.li.2020, yeplos\}@xjtu.edu.cn$^*$, liuf6@sustech.edu.cn$^\ddag$, \{yuanjun.shen, c.masouros\}@ucl.ac.uk$^\dag$}
}

\maketitle

\begin{abstract}
In this paper, we propose an interference exploitation symbol-level precoding (SLP) method for multi-level modulations via an in-block power allocation scheme to greatly reduce the signaling overhead. Existing SLP approaches require the symbol-level broadcast of the rescaling factor to the users for correct demodulation, which hinders the practical implementation of SLP. The proposed approach allows a block-level broadcast of the rescaling factor as done in traditional block-level precoding, greatly reducing the signaling overhead for SLP without sacrificing the performance. Our derivations further show that the proposed in-block power allocation enjoys an exact closed-form solution and thus does not increase the complexity at the base station (BS). In addition to the significant alleviation of the signaling overhead validated by the effective throughput result, numerical results demonstrate that the proposed power allocation approach also improves the error-rate performance of the existing SLP. Accordingly, the proposed approach enables the practical use of SLP in multi-level modulations.
\end{abstract}

\begin{IEEEkeywords}
MIMO, symbol-level precoding, constructive interference, multi-level modulations, signaling overhead.
\end{IEEEkeywords}

\section{Introduction}
Precoding design has been extensively studied for multi-user transmission in the field of multi-antenna communication systems \cite{intro-1}. By exploiting the channel state information (CSI), linear precoding methods have been studied in the literature \cite{intro-2}, \cite{intro-4}, which however do not exploit the information of the data symbols that is also available at the base station (BS) before transmission. Recently, interference exploitation symbol-level precoding (SLP) that employs the information of both the data symbols to be transmitted and the channel knowledge has emerged as a new precoding method \cite{ci-tut}, \cite{ci-tut2}. Compared to common block-level precoding where the precoding matrix is applied to a block of data symbols, SLP applies different precoding matrix to different data symbols or directly designs the precoded signals to be transmitted at the antenna port, which allows to observe interference that is inherent in multi-user transmission on a symbol-by-symbol basis and further exploit it instead of cancelling it, thus greatly enhancing the performance of multi-antenna communication systems. Due to such benefits, SLP has gained increasing research attention in recent years.

The study of interference exploitation precoding originates from the adaptation of traditional zero-forcing (ZF) and regularized ZF (RZF) precoding to closed-form SLP, where the concept of constructive interference (CI) and destructive interference (DI) is characterized \cite{ci-1}\nocite{ci-2}-\cite{ci-mag}. More specifically, \cite{ci-1} retains the CI with a dynamic linear precoding scheme while fully removing the DI via ZF, while a more superior scheme is proposed in \cite{ci-2} that further rotates the phases of DI such that all the interfering signals are constructive to the users. As a step further,  optimization-based CI precoding is studied in \cite{ci-3} for PSK signaling under the context of vector perturbation precoding, where CI in the form of symbol scaling is proposed. In \cite{ci-4}\nocite{ci-5}-\cite{ci-6}, CI precoding for signal-to-interference-plus-noise ratio (SINR)-constrained power minimization problems with PSK signaling is studied based on phase rotation, where the superiority of non-strict phase rotation is presented over the strict phase rotation considered in \cite{ci-1}, \cite{ci-2}, and \cite{ci-7} has further extended the CI precoding to QAM modulations. A distance-preserving CI metric is introduced in \cite{ci-10}, which considers the power-constrained SINR balancing problems. Importantly, \cite{ci-8} and \cite{ci-9} have further derived the optimal precoding structure for CI precoding with PSK signaling and QAM signaling respectively, which greatly reduces the computational costs of CI precoding. While the above works have shown to be effective in addressing the complexity issue of SLP, it should be noted that the above SLP solutions for power-constrained SINR balancing problems with QAM modulations have assumed that the rescaling factor is known to the users, which would however result in an excessive signaling overhead and hinder the practical implementation of CI precoding, since the rescaling factor needs to be broadcast to the users on a symbol level for these power-constrained SINR balancing SLP methods.

Therefore in this paper, we aim to reduce such excessive signaling overhead for SLP and propose a practical SLP scheme for multi-level modulations via an in-block power allocation. Specifically, as opposed to traditional SLP schemes that consider a uniform power allocation for each symbol duration within a transmission block, we propose to allocate the available transmit power of SLP dynamically for each symbol duration within a transmission block, based on which a two-stage optimization problem on the precoding matrix and the power allocation strategy is formulated. We prove via the Karush-Kuhn-Tucker (KKT) conditions that the proposed dynamic power allocation will lead to an identical rescaling factor within a transmission block if we firstly design the precoding matrix followed by the design of the power allocation strategy, which then allows the block-level broadcast of the rescaling factor to the users as usually done in traditional block-level precoding and greatly reduces the signaling overhead for SLP. Our derivations further reveal an exact closed-form solution for the proposed in-block power allocation strategy, leading to only negligible complexity increase at the BS. Numerical results have validated the significant signaling overhead reduction brought by the proposed in-block power allocation scheme without sacrificing the error-rate performance of SLP, which further boosts the practical deployment of SLP for multi-level modulations.

{\it Notations}: $a$, $\bf a$, and $\bf A$ denote scalar, column vector and matrix, respectively. ${( \cdot )^\text{T}}$, ${( \cdot )^\text{H}}$, ${( \cdot )^{-1}}$ denote transposition, conjugate transposition and inverse, respectively. ${{\cal C}^{n \times n}}$ and ${{\cal R}^{n \times n}}$ represent the sets of $n\times n$ complex- and real-valued matrices, respectively. $\Re \left\{ \cdot \right\}$ and $\Im \left\{ \cdot \right\}$ extract the real and imaginary part, $\circ$ is the Hadamard product, $\left\|  \cdot  \right\|_2$ represents the $\ell_2$-norm and $\jmath$ is the imaginary unit.

\section{System Model and Constructive Interference}
\subsection{System Model}
We focus on a generic multi-user multiple-input single-output (MU-MISO) communication system in the downlink, where the BS with $N_\text{T}$ transmit antennas communicates with a total number of $K$ single-antenna users simultaneously in the same time-frequency resource, where $K \le N_\text{T}$. Assume that a transmission block consists of $M$ symbol durations, within which the wireless channel stays constant, and we can express the data symbol matrix as
\begin{equation}
{\bf S}=\left[ { {\bf s}^{\left( {1}\right)}, {\bf s}^{\left( {2}\right)}, \cdots, {\bf s}^{\left( {M}\right)} } \right] \in {\cal C}^{K \times M}, 
\label{eq_1}
\end{equation}
where ${\bf s}^{\left( {m}\right)}=\left[ {s_1^{\left( m \right)}, s_2^{\left( m \right)}, \cdots, s_K^{\left( m \right)}} \right]^\text{T} \in {\cal C}^{K \times 1}$ is the data symbol vector in the $m$-th symbol duration drawn from a nominal QAM constellation, which is a representative example for multi-level modulations. Accordingly, the received signal for user $k$ in the $m$-th symbol duration can be expressed as
\begin{equation}
y_k^{\left( {m}\right)}=\sqrt{p^{\left( {m}\right)}} \cdot {\bf h}_k^\text{T} {\bf W}^{\left( {m}\right)} {\bf s}^{\left( {m}\right)} + n_k^{\left( {m}\right)},
\label{eq_2}
\end{equation}
where ${y}_k^{\left( {m}\right)}$ is the received signal for user $k$ in the $m$-th symbol duration, $p^{\left( {m}\right)}$ represents the allocated transmit power for the $m$-th symbol duration, ${\bf h}_k^\text{T} \in {\cal C}^{1 \times N_\text{T}}$ is the flat-fading Rayleigh channel between the BS and the users that is constant within the entire transmission block, ${\bf W}^{\left( {m}\right)} \in {\cal C}^{N_\text{T} \times K}$ denotes the precoding matrix, and $n_k^{\left( {m}\right)}$ is the corresponding additive Gaussian noise at the users with zero mean and variance $\sigma^2$. Since we focus on SLP and have included the allocated transmit power $p^{\left( {m}\right)}$ in \eqref{eq_2}, the symbol-level power constraint is enforced as $\left \| \mathbf{W}^{\left ( m \right )} \mathbf{s}^{\left ( m \right )} \right \|_{2}^2 = 1$, and we have $\sum_{m=1}^{M} p^{\left ( m \right )} \leq P_{\text{T}}$, where $P_\text{T}$ represents the total available transmit power for the transmission block. 

At the receiver side, $y_k^{\left( {m}\right)}$ needs to be scaled for correct demodulation when multi-level modulations are employed, and the signals ready for demodulation can be expressed as
\begin{equation}
r_k^{\left( {m}\right)}= f^{\left( {m}\right)} y_k^{\left( {m}\right)}= f^{\left( {m}\right)} \sqrt{p^{\left( {m}\right)}} \cdot {\bf h}_k^\text{T} {\bf W}^{\left( {m}\right)} {\bf s}^{\left( {m}\right)} + f^{\left( {m}\right)} n_k^{\left( {m}\right)},
\label{eq_3}
\end{equation}
where $f^{\left( {m}\right)}$ is the rescaling factor for the $m$-th symbol duration, also termed as the noise amplification factor \cite{ci-9}. Since ${\bf W}^{\left( {m}\right)}$ and ${\bf s}^{\left( {m}\right)}$ are known, $f^{\left( {m}\right)}$ can be calculated at the transmitter side but is not known at the receiver side, and therefore needs to be broadcast to each user on a symbol level for SLP.

\begin{figure}[!t]
\centering
\includegraphics[width=0.3\textwidth]{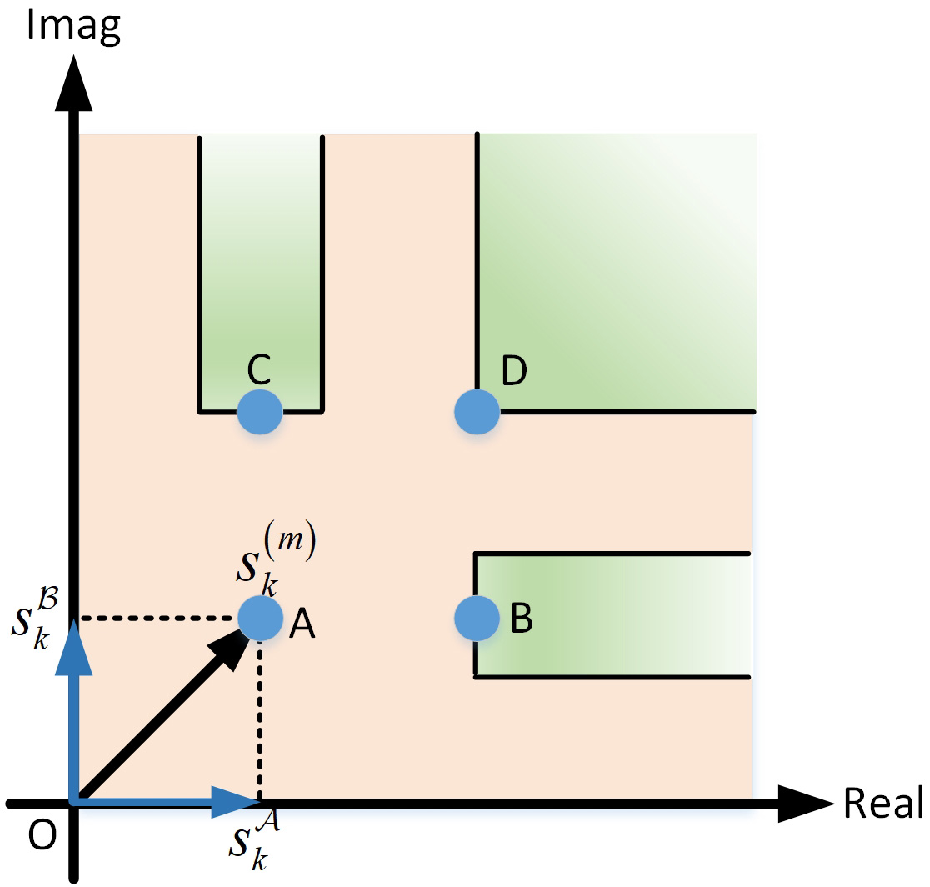}
\caption{One quarter of a 16QAM constellation}
\end{figure}

\subsection{Constructive Interference}
CI can be defined as the interference that pushes the received signal of interest further away from all of the decision boundaries \cite{ci-tut}, which effectively increases the distance between the received signal and the decision boundaries and further improves the performance of demodulation. As an illustrative example, we depict one quarter of a 16QAM constellation in Fig. 1, where the green shaded area represents the constructive region, within which the received signal receives CI, while the residual area in the QAM constellation is the destructive region, within which the corresponding signal receives DI. Based on the definition of CI, for QAM modulations CI can be exploited by the outermost constellation points belonging to type B, C and D in Fig. 1 (which will be explained later in Section III-A), while all the interference is seen as destructive for the inner constellation points. 

\section{Proposed Two-Stage Solution}
In this section, we introduce the proposed two-stage optimization on the precoding ${\bf W}^{\left( {m}\right)}$ and power allocation ${p}^{\left( {m}\right)}$, where the optimization on ${\bf W}^{\left( {m}\right)}$ is firstly designed, based on which the optimization on ${p}^{\left( {m}\right)}$ is performed.

\subsection{Optimization on ${\bf W}^{\left( {m}\right)}$}
In traditional SLP approach, uniform transmit power is assumed for each symbol duration, i.e., $p^{\left( m \right)}=\frac{P_\text{T}}{M}$, $\forall m \in {\cal M}$, where ${\cal M} = \left\{ {1,2,\cdots,M} \right\}$. In this letter, we aim to optimize both the transmit power and the precoding matrix for each symbol duration such that the minimum CI effect within the considered transmission block is maximized, and such a two-stage optimization will return an identical rescaling factor within the transmission block, as will be mathematically shown in Section III-B. Following \cite{ci-9}, we divide the constellation points for a QAM modulation into 4 types, as shown in Fig. 1. We follow the symbol-scaling CI metric in \cite{ci-9} and decompose each constellation point into
\begin{equation}
s_k^{\left( m \right)}=s_k^{\cal A} + s_k^{\cal B},
\label{eq_4}
\end{equation}
where $s_k^{\cal A}=\Re \left\{ s_k^{\left( m \right)} \right\} $ and $s_k^{\cal B} = \jmath \cdot  \Im \left\{ s_k^{\left( m \right)} \right\} $. Following a similar principle, we decompose the noiseless received signal ${\bf h}_k^\text{T} {\bf W}^{\left( {m}\right)} {\bf s}^{\left( {m}\right)}$ into
\begin{equation}
{\bf h}_k^\text{T} {\bf W}^{\left( {m}\right)} {\bf s}^{\left( {m}\right)}=\alpha_k^{\cal A} s_k^{\cal A} + \alpha_k^{\cal B}s_k^{\cal B},
\label{eq_5}
\end{equation}
where $\alpha_k^{\cal A}$ and $\alpha_k^{\cal B}$ are two introduced real scalars that jointly determine the effect of interference on $s_k^{\left( m \right)}$, and for simplicity of notation we have removed the index of symbol duration for $\alpha_k^{\cal A}$, $\alpha_k^{\cal B}$, $s_k^{\cal A}$ and $s_k^{\cal B}$. 

Based on Fig. 1, we observe that for QAM modulations, all the interference for 1) the real and imaginary part of constellation point type `A', 2) the real part of constellation point type `C', and 3) the imaginary part of constellation point type `B' is destructive. Therefore, to guarantee the correct demodulation of these constellation points after precoding, the relative position for constellation type `A', the real part of constellation type `C' and the imaginary part of constellation type `B' should remain constant, i.e., the following conditions need to be satisfied:
\begin{equation}
\alpha_p^{\cal A}=\alpha_p^{\cal B}, {\kern 3pt} \alpha_p^{\cal A}=\alpha_q^{\cal A}, {\kern 3pt} \alpha_p^{\cal B}=\alpha_j^{\cal B},
\end{equation}
if $s_p$ belongs to constellation type `A', $s_q$ belongs to constellation type `C', and $s_j$ belongs to constellation type `B'. For notational simplicity, we further define a set $\cal I$ that consists of the scaling factors for these inner constellation points, given by
\begin{equation}
{\cal I}=\left\{ {\alpha_p^{\cal A}, \alpha_p^{\cal B}, \alpha_q^{\cal A}, \alpha_j^{\cal B}, \forall p,q,j} \right\}=\left\{ {\alpha_n^{\cal I}, \forall n} \right\},
\end{equation}
and we can obtain that each $\alpha_n^{\cal I}$ should be identical for correct demodulation. Similarly, we can also define a set $\cal O$ consisting of the outer constellation points that can exploit CI, i.e., 1) the real part of constellation point type ‘B’, 2) the imaginary part of constellation point type ‘C’, and 3) the real and imaginary part of constellation point type ‘D’, given by
\begin{equation}
{\cal O}=\left\{ {\alpha_p^{\cal A}, \alpha_p^{\cal B}, \alpha_q^{\cal A}, \alpha_j^{\cal B}, \forall p,q,j} \right\}=\left\{ {\alpha_l^{\cal O}, \forall l} \right\},
\end{equation}
if $s_p$ belongs to constellation type `D', $s_q$ belongs to constellation type `B', and $s_j$ belongs to constellation type `C', and we obtain that to achieve CI, the value of the scaling factors in $\cal O$ should be made not smaller than the value of the scaling factors in $\cal I$.

Accordingly, the optimization problem on ${\bf W}^{\left( {m}\right)}$ to maximize the CI effect for the $m$-th symbol duration can then be formulated as \cite{ci-9}
\begin{equation}
\begin{aligned}
&\mathcal{P}_1: {\kern 3pt} \mathop {\max }\limits_{{\bf W}^{\left( {m}\right)}, t^{\left( {m}\right)}, \alpha_k^{\cal A}, \alpha_k^{\cal B}} {\kern 5pt} t^{\left( {m}\right)} \\
&{\kern 0pt} s. t. {\kern 10pt} {\bf C1:} {\kern 3pt} {\bf h}_k^\text{T} {\bf W}^{\left( {m}\right)} {\bf s}^{\left( {m}\right)}=\alpha_k^{\cal A} s_k^{\cal A} + \alpha_k^{\cal B}s_k^{\cal B}, {\kern 3pt} \forall k \in {\cal K}; \\
&{\kern 24pt} {\bf C2:} {\kern 3pt} t^{\left( {m}\right)} \le \alpha_l^{\cal O}, {\kern 3pt} \forall \alpha_l^{\cal O} \in {\cal O};\\
&{\kern 24pt} {\bf C3:} {\kern 3pt} t^{\left( {m}\right)} = \alpha_n^{\cal I}, {\kern 3pt} \forall \alpha_n^{\cal I} \in {\cal I};\\
&{\kern 24pt} {\bf C4:} {\kern 1pt} \left \| \mathbf{W}^{\left ( m \right )} \mathbf{s}^{\left ( m \right )} \right \|_{2}^2 = 1,
\label{eq_6}
\end{aligned}
\end{equation}
where ${\cal K}=\left\{ {1,2,\cdots,K} \right\}$. Although ${\cal P}_1$ is not convex due to the strict equality power constraint in ${\bf C4}$, we note that ${\cal P}_1$ can be transformed into a convex problem by relaxing the equality constraint in {\bf C4} into inequality, given by
\begin{equation}
\begin{aligned}
&\mathcal{P}_2: {\kern 3pt} \mathop {\max }\limits_{{\bf W}^{\left( {m}\right)}, t^{\left( {m}\right)}, \alpha_k^{\cal A}, \alpha_k^{\cal B}} {\kern 5pt} t^{\left( {m}\right)} \\
&{\kern 0pt} s. t. {\kern 10pt} {\bf C1:} {\kern 3pt} {\bf h}_k^\text{T} {\bf W}^{\left( {m}\right)} {\bf s}^{\left( {m}\right)}=\alpha_k^{\cal A} s_k^{\cal A} + \alpha_k^{\cal B}s_k^{\cal B}, {\kern 3pt} \forall k \in {\cal K}; \\
&{\kern 24pt} {\bf C2:} {\kern 3pt} t^{\left( {m}\right)} \le \alpha_l^{\cal O}, {\kern 3pt} \forall \alpha_l^{\cal O} \in {\cal O};\\
&{\kern 24pt} {\bf C3:} {\kern 3pt} t^{\left( {m}\right)} = \alpha_n^{\cal I}, {\kern 3pt} \forall \alpha_n^{\cal I} \in {\cal I};\\
&{\kern 24pt} {\bf C4:} {\kern 1pt} \left \| \mathbf{W}^{\left ( m \right )} \mathbf{s}^{\left ( m \right )} \right \|_{2}^2 \le 1,
\label{eq_6}
\end{aligned}
\end{equation}
which shares the same optimal solution as ${\cal P}_1$ since the strict equality in ${\bf C4}$ is always achieved for the optimality of ${\cal P}_2$, as shown in \cite{ci-9}. Moreover, we observe that the constraint ${\bf C3}$ indicates that the nominal constellation is scaled by $t^{\left( {m}\right)}$ due to the effect of the wireless channel, and together with \eqref{eq_2}, the rescaling factor for the received signals in the $m$-th symbol duration can be obtained as: 
\begin{equation}
f^{\left( {m}\right)}=\frac{1}{t^{\left( {m}\right)}\sqrt{p^{\left( {m}\right)}}}.
\label{eq_7}
\end{equation}

\subsection{Proposed In-Block Power Allocation}
Given that the optimization on ${\bf W}^{\left( {m}\right)}$ in ${\cal P}_1$ has been well studied in \cite{ci-9}, in the following we focus on the optimization on $p^{\left( {m}\right)}$ for given $t^{\left( {m}\right)}$. Accordingly, the optimization on $p^{\left( {m}\right)}$ to maximize the minimum CI effect within a transmission block, which is equivalent to minimizing the maximum noise amplification effect $f^{\left( {m}\right)}$, can be constructed as:
\begin{equation}
\begin{aligned}
&\mathcal{P}_3: {\kern 3pt} \mathop {\max } \limits_{p^{\left( {m}\right)}} \mathop {\min } \limits_{m} {\kern 3pt} t^{\left( {m}\right)}\sqrt{p^{\left( {m}\right)}} \\
&{\kern 0pt} s. t. {\kern 10pt} {\bf C1:} {\kern 3pt} \sum_{m=1}^{M} p^{\left ( m \right )} \leq P_{\text{T}}.
\label{eq_9}
\end{aligned}
\end{equation}
By introducing an auxiliary variable $u_m=\sqrt{p^{\left ( m \right )}}$, the above problem can be further expressed in a standard convex form:
\begin{equation}
\begin{aligned}
&\mathcal{P}_4: {\kern 3pt} \mathop {\min } \limits_{u_m} {\kern 3pt} -g \\
&{\kern 0pt} s. t. {\kern 10pt} {\bf C1:} {\kern 3pt} g - t^{\left( {m}\right)}{u_m} \le 0, {\kern 3pt} \forall m \in {\cal M} ; \\
&{\kern 23.5pt} {\bf C2:}\sum_{m=1}^{M} u_m^2 - P_{\text{T}} \le 0,
\label{eq_10}
\end{aligned}
\end{equation}
based on which the following proposition is obtained.

{\bf Proposition 1:} When the optimality of ${\cal P}_4$ is achieved, we arrive at an identical rescaling factor for each symbol duration in the transmission block, i.e, $f^{\left( {1}\right)}=f^{\left( {2}\right)}=\cdots=f^{\left( {M}\right)}$.

{\bf Proof:} We prove this proposition via the KKT conditions, where the Lagrangian of ${\cal P}_4$ is formulated as
\begin{equation}
\begin{aligned}
&{\cal L} \left( {u_m ,g, \delta_m, \vartheta } \right)= -g+\sum_{m=1}^{M}\delta_m\left ( g-t^{\left ( m \right )} u_{m} \right ) \\
& {\kern 77pt} + \vartheta \left( {\sum_{m=1}^{M} u_m^2 - P_{\text{T}}} \right) \\
& {\kern 20pt} = \left( {{\bf 1}^\text{T} {\bm \delta} -1} \right)g + \vartheta \cdot {\bf u}^\text{T} {\bf u} - \left( {{\bm \delta} \circ {\bf t}} \right)^\text{T} {\bf u} - \vartheta \cdot P_{\text{T}}.
\end{aligned}
\label{eq_11}
\end{equation}
In \eqref{eq_11}, ${\bm \delta}=\left[ {\delta_1,\delta_2,\cdots,\delta_M} \right]^\text{T}$, ${\bf u}$ and ${\bf t}$ are similarly defined, and ${\bf 1}=\left[ {1,1,\cdots,1} \right]^\text{T}$.

Based on \eqref{eq_11}, the corresponding KKT conditions for ${\cal P}_4$ are given by
\begin{IEEEeqnarray}{rCl} 
\IEEEyesnumber
\frac{{\partial {\cal L}}}{{\partial g}} =  \left( {{\bf 1}^\text{T} {\bm \delta} -1} \right)  = 0 {\kern 10pt} \IEEEyessubnumber* \label{eq_12a} \\
\frac{{\partial {\cal L}}}{{\partial {\bf u} }} = \vartheta \cdot {\bf u} - \left( {{\bm \delta} \circ {\bf t}} \right) = {\bf{0}}, {\kern 2pt} \forall m \in {\cal M} {\kern 10pt} \label{eq_12b} \\
\delta_m\left ( g-t^{\left ( m \right )} u_{m} \right ) = 0, {\kern 2pt} \forall m \in {\cal M} {\kern 10pt} \label{eq_12c} \\
\vartheta \left( {\sum_{m=1}^{M} u_m^2 - P_{\text{T}}} \right) = 0 {\kern 10pt} \label{eq_12d}
\end{IEEEeqnarray}
where $\delta_m$, $\forall m \in {\cal M}$ and $\vartheta$ are the introduced Lagrange multipliers corresponding to the inequality constraints {\bf C1} and {\bf C2} in ${\cal P}_4$, respectively. By observing (12), firstly we obtain that  $\vartheta \ne 0$, since $\vartheta = 0$ results in $\delta_m=0$, $\forall m \in {\cal M}$ based on \eqref{eq_12b}, which contradicts with \eqref{eq_12a}. This means that the transmit power constraint is active when optimality is achieved, i.e., $\sum_{m=1}^{M} u_m^2 = P_{\text{T}}$. Subsequently, according to \eqref{eq_12c} and given that $\delta_m \ne 0$, $\forall m \in {\cal M}$, we further obtain $g-t^{\left ( m \right )} u_{m}=0$, $\forall m \in {\cal M}$, which is equivalent to
\begin{equation}
t^{\left( {1}\right)}{u_1}=t^{\left( {2}\right)}{u_2}=\cdots=t^{\left( {M}\right)}{u_M},
\label{eq_13}
\end{equation}
and with $u_m=\sqrt{p^{\left ( m \right )}}$, we arrive at
\begin{equation}
t^{\left( {1}\right)}\sqrt{p^{\left ( 1 \right )}}=t^{\left( {2}\right)}\sqrt{p^{\left ( 2 \right )}}=\cdots=t^{\left( {M}\right)}\sqrt{p^{\left ( M \right )}},
\label{eq_14}
\end{equation}
which based on \eqref{eq_7} completes the proof. $\blacksquare$

Moreover, we can further derive the closed-form solution of the optimal transmit power value $p^{\left ( m \right )}$, given by the following proposition.

{\bf Proposition 2:} The value of the optimal transmit power $p^{\left ( m \right )}$ can be obtained in a closed form as
\begin{equation}
p^{\left ( m \right )}=\frac{\frac{1}{\left( t^{\left ( m \right )} \right)^2}}{\sum_{m=1}^{M} \frac{1}{\left( t^{\left ( m \right )} \right)^2} } \cdot P_\text{T},
\label{eq_15}
\end{equation}
and the value of the rescaling factor in the considered transmission block is given by
\begin{equation}
f^{\left( {1}\right)}=f^{\left( {2}\right)}=\cdots=f^{\left( {M}\right)}=\sqrt{\frac{\sum_{m=1}^{M} \frac{1}{\left( t^{\left ( m \right )} \right)^2} }{P_\text{T}}}.
\label{eq_f}
\end{equation}

{\bf Proof:} The closed-form solution can readily be obtained based on the results in {\bf Proposition 1} where $\sum_{m=1}^{M} p^{\left ( m \right )} = P_{\text{T}}$ and $t^{\left( {1}\right)}\sqrt{p^{\left ( 1 \right )}}=t^{\left( {2}\right)}\sqrt{p^{\left ( 2 \right )}}=\cdots=t^{\left( {M}\right)}\sqrt{p^{\left ( M \right )}}$, and the value of $f^{\left( {m}\right)}$ can be obtained accordingly. $\blacksquare$

Based on {\bf Proposition 2} above, it is observed that the rescaling factor for each symbol duration in the considered transmission block is identical and therefore only needs to be fed to the receivers once in a transmission block, which greatly reduces the signaling overhead for SLP with multi-level constellations.

\section{Numerical Results}
In this section, we evaluate the performance of SLP with the proposed in-block power allocation via Monte Carlo simulations. We consider a practical communication system where the BS employs $B$ bits to broadcast the rescaling factor $f$ to the users, which is subject to quantization errors due to the limited feed-forwarding, given by
\begin{equation}
\hat f= f + \epsilon ,
\label{eq_16}
\end{equation}
where $\hat f$ is the rescaling factor received at the users, and $f$ is the ideal rescaling factor known at the BS as in \eqref{eq_3}. $\epsilon \sim {\cal N} \left( {0, \nu } \right)$ is the quantization error, whose variance is modeled according to \cite{ci-3}, \cite{model-1}, \cite{model-2} as
\begin{equation}
\nu=\frac{f_\text{max}}{2^B},
\label{eq_17}
\end{equation}
where $f_\text{max}$ is the maximum value of the rescaling factor with non-zero probability. Accordingly, we evaluate the performance of different precoding techniques via the effective throughput for $\mathbb{M}$-QAM, defined as throughput minus the total number of bits required for feed forwarding, given by
\begin{equation}
\begin{aligned}
T_\text{eff} &=\text{max} \left\{ { \left( {1-\text{BLER}} \right) \cdot \log _{2}\left ( \mathbb{M} \right ) \cdot K - N_\text{overhead}, 0 } \right\} \\
&= \text{max} \left\{ { \left( {1-P_\text{b}} \right)^{M\cdot \log _{2}\left ( \mathbb{M} \right )} \cdot \log _{2}\left ( \mathbb{M} \right ) \cdot K-N_\text{overhead}, 0 } \right\},
\end{aligned}
\end{equation}
where BLER is the block error rate, and $P_\text{b}$ is the bit error rate (BER). $N_\text{overhead}$ is the total number of bits for signaling overhead, and $N_\text{overhead}=\frac{B}{M}$ for ZF, RZF, and SLP with in-block power allocation, while $N_\text{overhead}=B$ for traditional SLP with uniform power allocation. Without loss of generality, we assume that the number of bits for broadcasting the rescaling factor to the users is $B=5$ bits. The maximum available transmit power in a transmission block is $P_\text{T}=1$, and the transmit SNR in each symbol duration is accordingly defined as $\rho=\frac{1}{M \sigma^2}$. We compare our proposed scheme with traditional block-level ZF and RZF precoding via 5000 channel realizations, and both 16QAM and 64QAM modulations are considered in the simulations.


\begin{figure*}[!t]
\captionsetup[subfigure]{labelformat=empty}
\captionsetup{labelformat=empty}
\begin{centering}
\subfloat[Fig. 2  Value of $f^{\left( {m}\right)}$, $M=10$, $\text{SNR}=40$dB]
{\begin{centering}
\includegraphics[width=0.3\textwidth]{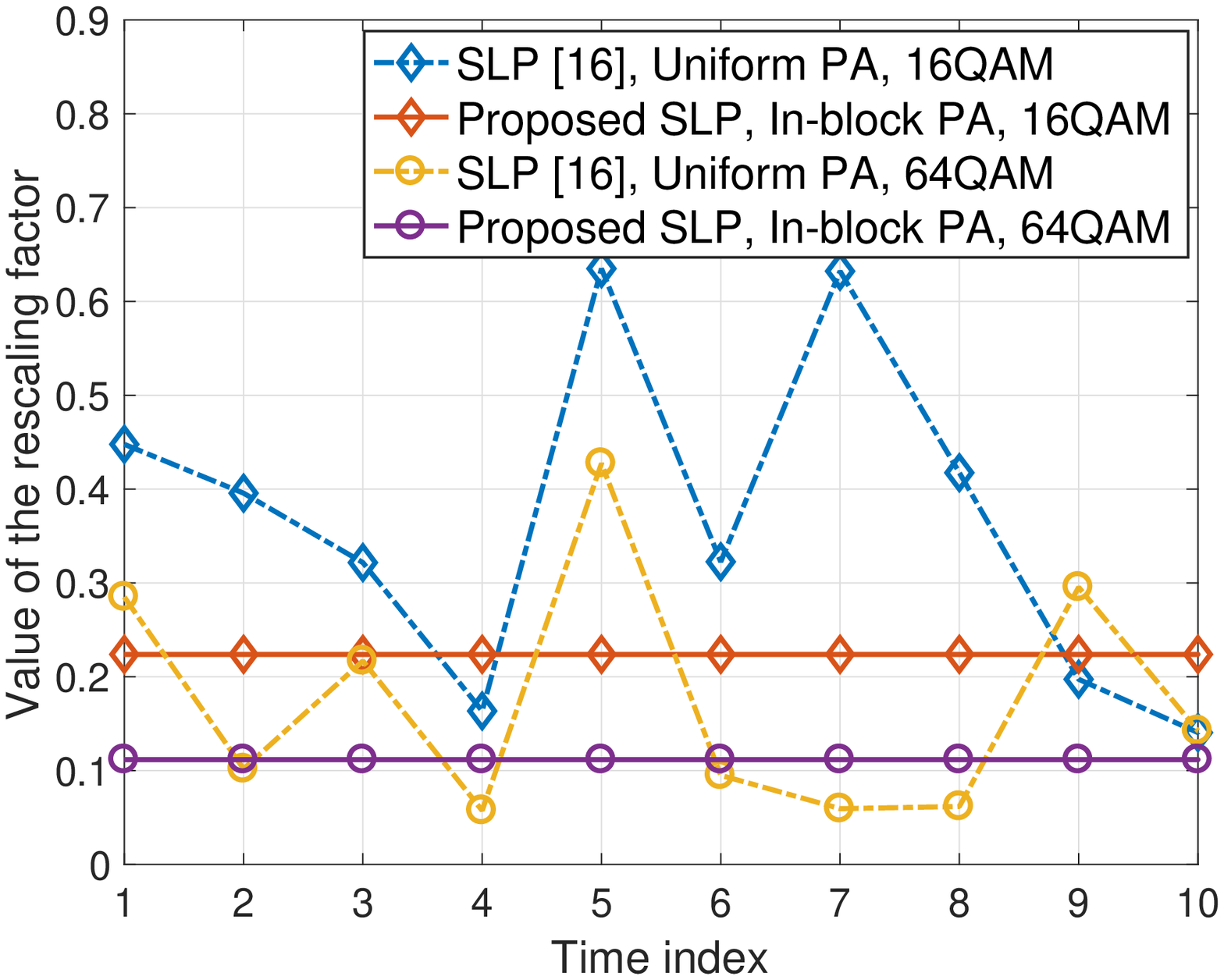}
\par
\end{centering}
}
\hspace{0.1cm}
\subfloat[Fig. 3  BER v.s. SNR, $M=200$]
{\begin{centering}
\includegraphics[width=0.3\textwidth]{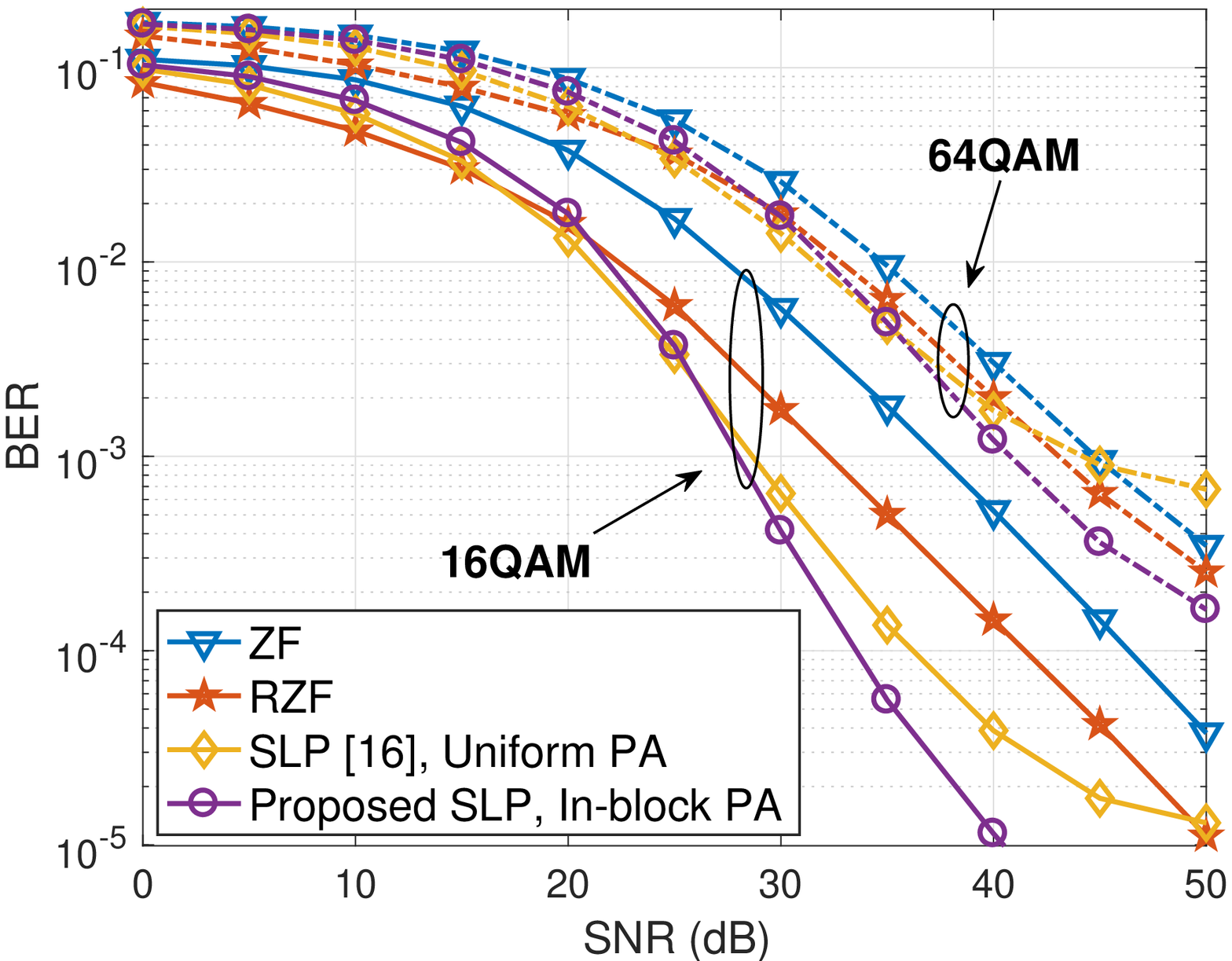}
\par
\end{centering}
}
\hspace{0.1cm}
\subfloat[Fig. 4  Effective throughput v.s. SNR, $M=200$]
{\begin{centering}
\includegraphics[width=0.3\textwidth]{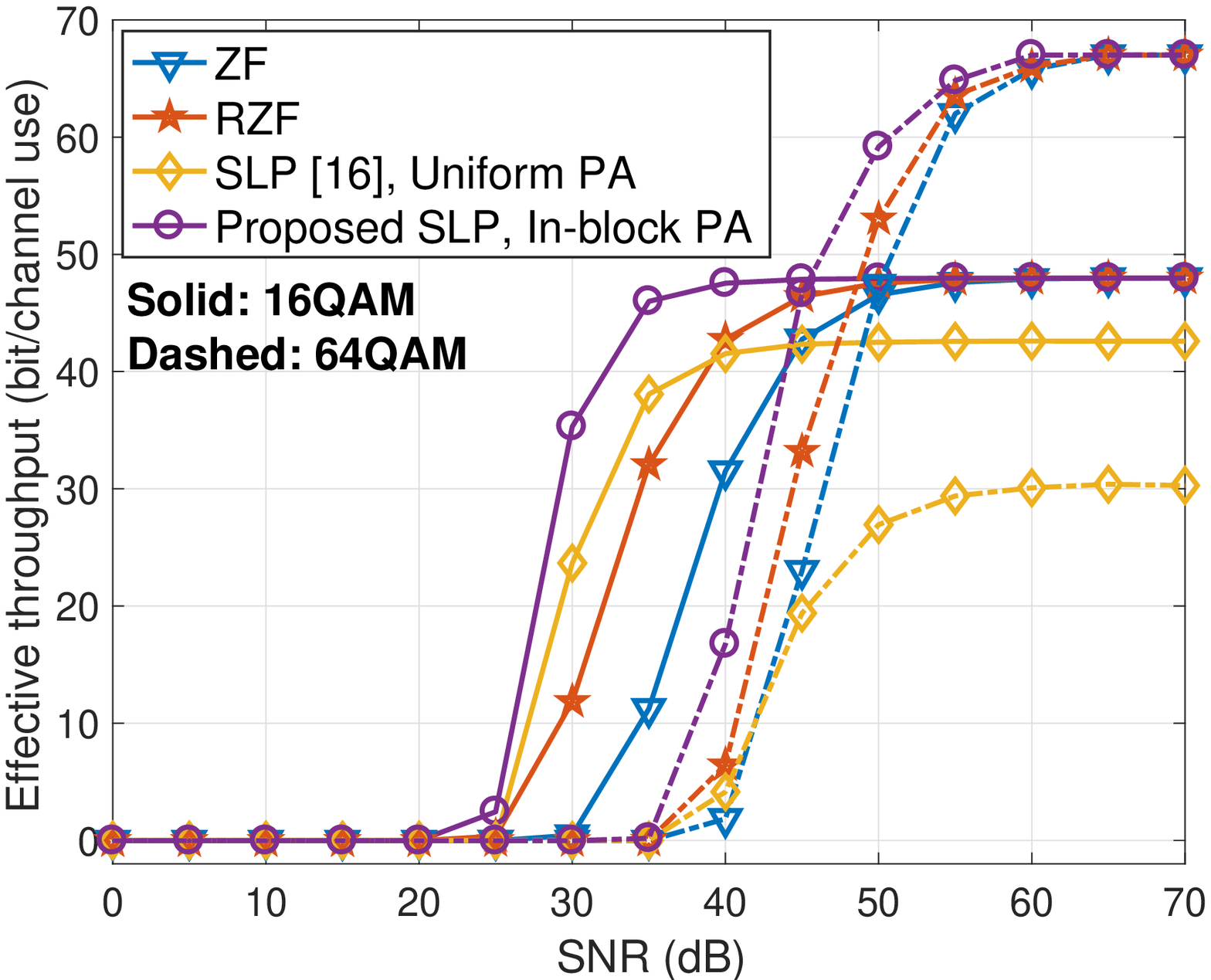}
\par
\end{centering}
}
\par
\end{centering}
\caption{Numerical results for 16QAM and 64QAM, $K=N_\text{T}=12$, $f_\text{max}=1$, $B=5$}
\end{figure*}

Fig. 2 validates the effectiveness of the proposed in-block power allocation in reducing the signaling overhead for both 16QAM and 64QAM, where we assume $M=10$ symbols for ease of illustration. Compared to traditional SLP approach where the value of the rescaling factor varies for each symbol duration in the transmission block, which requires a symbol-level broadcast of these values to the users for correct demodulation, SLP with the proposed in-block power allocation scheme returns an identical rescaling factor within a transmission block, which then enables the block-level broadcast of the rescaling factors as done in traditional block-level precoding, validating the effectiveness of the proposed scheme.

Fig. 3 compares the BER of SLP with the proposed in-block power allocation with ZF precoding with practical $M=200$ symbols, RZF precoding and SLP with uniform power allocation, for both 16QAM and 64QAM. Since we have considered quantization errors in broadcasting the rescaling factors to the users as in practical communication systems, the traditional SLP technique is observed to exhibit BER losses when the transmit SNR increases and becomes inferior to traditional ZF/RZF precoding. On the contrary, the SLP method with the proposed in-block power allocation has alleviated such performance degradation, and meanwhile still offers transmit SNR gains over traditional schemes.

Fig. 4 depicts the effective throughput for different precoding schemes to further highlight the significance of the proposed in-block power allocation strategy for SLP with practical $M=200$ symbols. As can be observed, traditional SLP methods have shown the worst effective throughput performance due to the requirement of symbol-level broadcast of the rescaling factor, which results in an excessive signaling overhead. Meanwhile, we observe that the proposed scheme can greatly improve the effective throughput for SLP and achieves the highest throughput performance over ZF and RZF precoding. Both of the BER and throughput results above have exhibited the superiority and significance of the proposed scheme for existing SLP with multi-level modulations.

\section{Conclusions}
In this letter, we have designed an in-block power allocation scheme for SLP with multi-level modulations, which returns an identical rescaling factor for the entire transmission block. The proposed scheme can thus greatly reduce the signaling overhead of existing SLP schemes by reducing the frequency for the broadcast of the rescaling factor from symbol level to block level, achieving a similar signaling overhead to traditional block-level precoding methods without sacrificing the performance. By proposing a two-stage precoding and power allocation design, our derivations have shown that the in-block power allocation enjoys a closed-form solution, which is thus efficient to deploy. In addition, numerical results demonstrate that the proposed power allocation scheme exhibits significant performance improvements for SLP in terms of the effective throughput over existing SLP techniques without in-block power allocation.

\bibliographystyle{IEEEtran}
\bibliography{refs.bib}

\end{document}